\documentstyle[12pt,fleqn,epsfig]{article}
\begin{document}
\begin{center}
{\Huge 
Relationship among Phenotypic Plasticity, Genetic and Epigenetic Fluctuations, Robustness, and Evolovability;
Waddington's Legacy revisited under the Spirit of Einstein}
\end{center}

\begin{Large}
\begin{center}
Kunihiko Kaneko\footnotemark[1]\ \footnotemark[2]{}
\end{center}
\end{Large}
\begin{center}
\noindent
\footnotemark[1] Department of Pure and Applied Sciences, Univ. of
Tokyo, 3-8-1 Komaba, Meguro-ku, \\
Tokyo 153-8902, Japan\\
\footnotemark[2] ERATO Complex Systems Biology Project, JST, 3-8-1
Komaba, Meguro-ku, \\
Tokyo 153-8902, Japan \\
\end{center}

\begin{abstract}
Questions on possible relationship between phenotypic plasticity and evolvability, as well as that between robustness and evolution have been addressed over decades in the field of evolution-development. By introducing an evolutionary stability assumption on the distribution of phenotype and genotype, we establish quantitative relationships on plasticity, phenotypic fluctuations, and evolvability.  Derived are proportionality among plasticity as a responsiveness of phenotype against environmental change, variances of phenotype fluctuations of genetic and developmental origins, and evolution speed. Confirmation of the relationships is given by numerical experiments of a gene expression dynamics model with an evolving transcription network, whereas verifications by laboratory evolution experiments are also discussed. These results provide quantitative formulation on canalization and genetic assimilation, in terms of fluctuations of gene expression levels.
\end{abstract}

\section{Introduction}

\subsection{Questions on evolution}

Q1:
How is evolvability characterized? Is it related with developmental process? Naively speaking, one may expect that if developmental process is rigid, phenotype is not easily changed, so that evolution does not progress so fast.  Then one may expect that evolvability is tightly correlated with plasticity as a result of developmental process.  Here plasticity refers to the changeability of phenotype as a result of developmental dynamics(Callahan et al. 1997, West-Eberhard, 2003, Kirschner and Gerhart 2005, Pigliucci et a;. 2006). Hence, the question we address here is with regard to the correlation between plasticity and evolvability (Ancel and Fontana, 2002). With such correlation, we may formulate relationship between evolvability and development quantitatively.

Q2:  Besides plasticity, an important issue in evolution is robustness.  Robustness is defined as the ability to function against possible changes in the system (Wagner 2000, 2005,  Wagner et al., 1997, Barkai and Leibler, 1997, Alon et al., 2000, Ciliberti et al, 2007, Kaneko 2007). In any biological system, these changes have two distinct origins: 
genetic and epigenetic. The former concerns structural robustness of the phenotype, i.e., rigidity of phenotype against genetic changes produced by mutations. On the other hand, the latter concerns robustness against the stochasticity that can arise in environment or during developmental process, which includes fluctuation in initial states and stochasticity occurring during developmental dynamics or in the external environment. 

Now, there are several questions associated with robustness and evolution.  Is robustness increased or decreased through evolution? Schmalhausen proposed stabilizing evolution (Schmalhausen 1949), and Waddington proposed canalization (Waddington 1942,1957).  Both of them discussed the possibility that robustness increases through evolution.  On the other hand, the reverse question, whether the evolvability is decreased by the increase of robustness has also been discussed. Since the robustness increases rigidity in phenotype, the plasticity as well as the evolvability may be decreased accordingly.

Besides these robustness-evolution questions, another issue to be addressed is the relationship between the two types of robustness,
genetic and epigenetic.  Do the two types of robustness increase or decrease in correlation, through evolution? In fact, the epigenetic robustness concerns with development. Higher epigenetic robustness means developmental stability in producing a given phenotype, whereas genetic robustness is stability against mutational change, i.e., evolutionary stability.  Hence the question on a possible relationship between genetic and epigenetic robustness concerns with the issue on evolution-development relationship, that is an important topic in biology.

\subsection{Formulation of the problem in terms of stochastic dynamical systems}

Here we study the questions raised above, in terms of fluctuation.  
Let us first discuss epigenetic robustness, i.e., the degree of phenotype change as a result of developmental process. 
The {\sl developmental dynamics} to shape a phenotype generally involves stochasticity.
As has been recognized extensively, gene expression dynamics are noisy. Accordingly, phenotype as well as fitness, of isogenic organisms is distributed (Oosawa, 1975, Spudich and Koshland, 1976). Still, phenotype of an organism is often reproducible even under fluctuating environment or under molecular fluctuations. Therefore, phenotype that is concerned with fitness is expected to keep some robustness against such stochasticity in gene expression, i.e., robustness in `developmental' dynamics to noise.

When the phenotype is less robust to noise, this distribution is broader.  Hence, the variance of isogenic phenotypic fluctuation distribution, denoted as $V_{ip}$, gives an index for  robustness to noise in developmental dynamics(Kaneko 2006,2007).  For example, distribution of protein abundances over isogenic individual cells has been measured, by using fluorescence proteins.  The fluorescence level which gives an estimate of the protein concentration is measured either by flow cytometry or at a single cell level by microscopes (McAdams and Arkin 1997, Elowitz et al. 2002, Hasty et al, 2000, Furusawa et al. 2005, Krishna et al., 2005, Bar-Even et al,, 2004, Kaern et al.2005). 

Genetic robustness is also measured in terms of fluctuation.  Due to mutation to genes, the phenotype (fitness) is distributed.  Since even phenotype of isogenic individuals is distributed, the variance of phenotype distribution of heterogenic population includes both the contribution from phenotypic noise in isogenic individuals and from genetic variance.  To distinguish the two, we first obtain the average phenotype over isogenic individuals, and then compute the variance of this average phenotype over heterogenic population. Then this variance is due only to genetic heterogeneity. This variance is denoted as $V_g$ here. Then, the robustness to mutation is estimated by this variance.  If $V_g$ is smaller, genetic change gives little influence on the phenotype, implying larger genetic (or mutational) robustness.  

  The phenotypic plasticity, on the other hand, represents the degree of change in phenotype against variation in environment.  Quantitatively, we define the following ``response ratio" as a measure of plasticity. It is defined by the ratio of change in phenotype to environmental change.  For example, measure the (average) concentration of a specific protein ($X$).  We measure this concentration by (slightly) changing the concentration of an external signal molecule ($S$).  Then, response ratio $R$ is computed as the change in the protein concentration divided by the change in the signal concentration ($R=dX/dS$).  This gives a quantitative measure for the phenotypic plasticity.  In general, there can be a variety of ways to change the environmental condition.  In this case, one may also use the average of such response ratio over a variety of environmental changes as a measure of phenotypic plasticity.

By using these variances the questions we address are represented as follows(Visser et al., 2000,West-Eberhard 2003, Weinig 2000, Gibson and Wagner 2000, Kirschner and Gerhart 2005, Ancel and Fontana 2002):
(i) Does the evolution speed increase or decrease with robustness?--- {\sl relationship between evolution speed with $V_g$ or $V_{ip}$} 
(ii) Is epigenetic and genetic robustness related ? --- {\sl relationship between $V_g$ and $V_{ip}$}
(iii) Does robustness in(de)crease through evolution? --- {\sl change of $V_g$ or $V_{ip}$ through evolution}
(iv) How is phenotypic plasticity related with robustness and evolution speed? --- {\sl relationship between $R$ and $V_g$ and $V_{ip}$}.

To answer these questions, we have carried out evolution experiments by using bacteria (Sato et al, 2003) and numerical simulations on catalytic reaction (Kaneko and Furusawa 2006) or gene networks (Kaneko 2007,2008), while we have also developed evolutionary stability theory at a population level. Here we briefly review these results and discuss relevance to biological plasticity.

\section{ Macroscopic Approach I;  Fluctuation-response relationship}

One might suspect that isogenic phenotype fluctuations $V_{ip}$ are not related to evolution, since phenotypic change without genetic change is not transferred to the next generation. However, the degree of fluctuation is determined by the gene, and accordingly it is heritable. Hence, there may be a relationship between isogneic phenotypic fluctuation and evolution.  
In fact, we have found evolutionary fluctuation--response relationship in bacterial evolution in the laboratory, where the evolution speed is positively correlated with the variance of the isogenic phenotypic fluctuation(Sato et al.2003). This correlation or proportionality was confirmed in a simulation of the reaction network model(Kaneko and Furusawa 2006). The origin of proportionality between the isogenic phenotypic fluctuation and genetic evolution has been discussed in light of the fluctuation--response relationship in statistical physics(Einstein 1926, Kubo et al. 1985).
In the present section we briefly summarize these recent studies.

\subsection{Selection experiment by bacteria}

We carried out a selection experiment to increase the fluorescence in bacteria and checked a possible relationship between evolution speed and isogenic phenotypic fluctuation.
First, by attaching a random sequence to the N terminus of a wild-type
Green Fluorescent Protein (GFP) gene, protein with low fluorescence was generated.  The gene for this protein was introduced into E.coli,
as the initial generation for the evolution.
By applying random mutagenesis only to  the attached fragment in the gene, a mutant pool with a diversity of cells was prepared. Then, cells with highest fluorescence intensity are selected for the next generation. With this procedure, the (average) fluorescence level of selected cells increases by generations.  The evolution speed at each generation is computed as the difference of the fluorescence levels between the two generations.
Then, to observe the isogenic phenotypic fluctuation, we use a distribution of clone cells of the selected bacteria. The distribution of the fluorescence intensity of the clones was measured with the help of flow cytometry, from which the variance of fluorescence was measured.

From these data we plotted evolution speed divided by the synonymous mutation rate versus the fluorescence variance.  The data support strong correlation between the two, suggesting the proportionality between the two (Sato et al. 2003).

\subsection{Supports from numerical experiments}

To confirm this relationship more accurately, we have also numerically studied a model of reproducing cells consisting of catalytic reaction networks.  Here the reaction networks of cells were slightly changed to produce mutants,  and those networks with a higher concentration of a given, specific chemical component were selected. Again, the evolution speed was measured as the increase of the concentration at each generation, and the fluctuation as the variance of the concentration over identical networks.  From extensive numerical experiments, the proportionality between the two was clearly confirmed(Kaneko and Furusawa 2006).

\subsection{Fluctuation-response relationship}

To understand the proportionality, we first propose a general relationship between fluctuation and response. First, we refer to a measurable phenotype quantity (e.g. the concentration of a protein) as a "variable" $x$ of the system, while we assume the existence of "parameter(s)"  which controls the change of a variable. The parameter corresponding to $x$ is indicated as $a$. In the case of adaptation, this parameter is a quantity to specify the environmental condition, while in the case of evolution, the parameter is an index representing  genotype that governs the corresponding phenotype variable.

Even among individuals sharing the identical gene (the parameter) $a$, the variables $x$ are distributed.  Then a distribution $P(x;a)$ of the variable $x$ over cells is defined, for a given parameter $a$ (e.g., genotype). Here the "average" and "variance" of $x$ are defined with regards to the distribution $P(x;a)$ 
(see Fig.1 for schematic representation of the fluctuation-response relationship).

Consider the change in the parameter value  $a$ to $a \rightarrow a+\delta a$. Then, the proposed fluctuation-response relationship (Sato et al., 2003, Kaneko 2006) is give by

\begin{equation} \frac{<x>_{a+\Delta a}-<x>_{a}}{\Delta a} \propto <(\delta x)^2>,\end{equation}

\noindent
where $<x>_a$ and $<(\delta x)^2>=<(x-<x>)^2>$ are the average and variance of the variable $x$ for a given parameter value $a$, respectively. 

The above relationship is derived by assuming that the distribution is approximately Gaussian.  At $a=a_0$, we then the distribution is written by

\begin{equation} P(x; a)={\sl N}exp(-\frac{(x-X_0)^2}{2\alpha(a)}+v(x,a)).  \end{equation}
with ${\sl N_0}$ a normalization constant so that $\int
P(x:a)dx=1$. 

Here, $X_0$ is the peak value of the variable at $a=a_0$, where the term $v(x,a)$ gives a deviation from the Gaussian form when $a$ is changed from $a_0$, so that $v(a,x)$ can be expanded as $v(a,x)=C(a-a_0)(x-X_0)+...$, with $C$ as a constant, where $...$ is a higher order term in $(a-a_0)$ and $(x-X_0)$ which are neglected.

Assuming this distribution form, the change of the average
values $<x>$ against the change of the parameter from $a_0$ to $a_0
+\Delta a$ is straightforwardly computed, which leads to
\begin{equation} \frac{<x>_{a=a_0+\Delta a}-<x>_{a=a_0}}{\Delta a}=C\alpha (a_0+\Delta a). \end{equation}

Noting that $\alpha = <(\delta x)^2>$ and neglecting deviation between $\alpha (a_0+\Delta a)$ and $\alpha (a_0) $, we get
\begin{equation} \frac{<x>_{a=a_0+\Delta a}-<x>_{a=a_0}}{\Delta a}=C
<(\delta x)^2>. \end{equation}
This relationship is a general result which holds for Gaussian-like distributions if higher-order changes in the distribution is neglected and if changes in parameters are represented by a  ``linear coupling term"  which brings about a shift of the average of the corresponding variable.

Although our formula is formally similar to that used in statistical physics, ours is approximate; the distribution must be close to Gaussian,  or the variable is scaled  so that the corresponding distribution is Gaussian.  Also, in general, the coefficient $C$ could depend on the parameter $a$, if the change in $a$ from $a_0$ is not small.

\begin{figure}[tbp]
\begin{center}
\includegraphics[width=12cm,height=8cm]{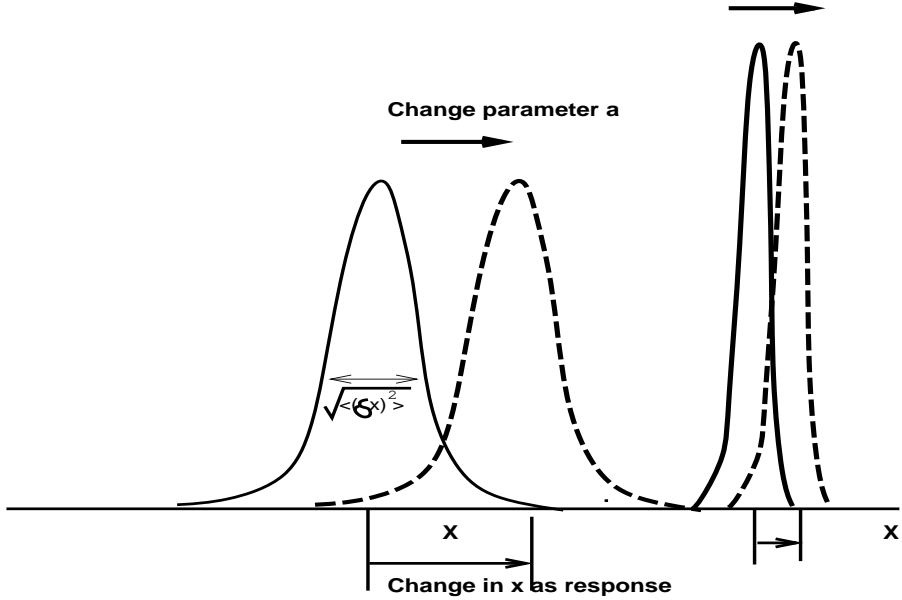}
\caption{
Schematic diagram on fluctuation-response relationship.
The distribution function $P(x;a)$ is shifted by the change of the parameter $a$.
If the variance is larger the shift is larger.}
\end{center} 
\end{figure}

\subsection{Plasticity-fluctuation relationship}

First we apply the above relation to plasticity.  Here we make the following correspondence:

(i) $a$ --- parameter to characterize the environmental condition

(ii) change in $a$ --- change in environment

(iii) $<x>$ --- variable characterizing a phenotypic state

(iv)change in $<x>$ ---change in the average phenotype due to the environmental change

(v) $<(\delta x)^2>$ --- variance of the phenotypic distribution at a given fixed environmental condition. 

Then, the relationship eq.(1) means that the plasticity is proportional to the fluctuation.  

\subsection{Evolutionary fluctuation-response relationship}

This fluctuation-response relationship can be applied to evolution by making the following correspondence:

 (i) $a$ --- a parameter that specifies the genotype 

 (ii) change in $a$ --- genetic mutation

(iii) change in $<x>$ ---change in the average phenotype due to the genetic change,  and

(iv) $<(\delta x)^2>$ --- variance of the phenotypic distribution over clones $V_{ip}$. 

Then, the above relationship (1) suggest that {\bf the evolution speed of a phenotype (i.e., the change of the average phenotype per generation) divided by the mutation rate is proportional to $V_{ip}$, the variance of isogenic phenotypic fluctuation}.  This is what we have obtained in the experiment and simulations.

The above proposition can be interpreted as follows: As the degree of
the fluctuations of the phenotype decreases, the change of the phenotype
under a given rate of genetic change also decreases.  In other words,
the larger the phenotypic fluctuations are, the larger the evolution
speed is. In this sense, we have a theoretical basis to characterize the degree of evolvability.  

\section{Macroscopic Approach II:  Stability theory for Genotype-Phenotype Mapping}

The evolutionary fluctuation-response relationship mentioned above casts another mystery to be solved. There is an established relationship between evolutionary speed and phenotypic fluctuation. It is the so-called fundamental theorem of natural selection by Fisher(Fisher 1930) which states that evolution speed is proportional to the variance of phenotype due to genetic variation, which is the phenotypic variance as a result of the distribution of genes in a heterogenic population? It is defined as the variance of {\em average} phenotype, $V_g$ in the present paper. In contrast, the evolutionary fluctuation--response relationship, proposed above, concerns phenotypic fluctuation of isogenic individuals as denoted by $V_{ip}$. While $V_{ip}$ is defined as a variance over clones, i.e., individuals with the same genes, $V_g$ is a result of the heterogenic population distribution. Hence, the fluctuation--response relationship and the relationship concerning $V_g$ by Fisher's theorem are not identical.

If $V_{ip}$ and $V_g$ are proportional through an evolutionary course, the two relationships are consistent. Such proportionality, however, is not self-evident, as $V_{ip}$ is related to variation against the developmental noise and $V_g$ against the mutation. The relationship between the two, if it exists, postulates a constraint on genotype--phenotype mapping and may create a quantitative formulation of a relationship between development and evolution.

Here we formulate a phenomenological theory to support the relationship (see Fig.2). Recall that originally the phenotype is given as a function of gene, so that we considered the conditional distribution function $P(x;a)$.  However, the genotype distribution is influenced by phenotype through the selection process, as the fitness for selection is a function of phenotype. Now, consider gradual evolutionary process. Then, it could be possible to assume that some ``quasi-static" distribution on genotype and phenotype is reduced as a result of feedback process from phenotype to genotype (see Fig.2). Considering this point, we hypothesize that there is a two-variable distribution $P(x,a)$ both for the phenotype $x$ and genotype $a$. 

By using this distribution, $V_{ip}$, variance of $x$ of the distribution for given $a$, can be written as $V_{ip}(a)=\int (x-\overline{x(a)})^2 P(x,a)dx$, where $\overline{x(a)}$ is the average phenotype of a clonal population sharing the genotype $a$, namely $\overline{x(a)}=\int P(x,a)x dx$.
$V_g$ is defined as the variance of the average $\overline{x(a)}$, over genetically heterogeneous individuals and is given by $V_{g}=\int (\overline{x(a)}-<\overline{x}>)^2 p(a)da$, where $p(a)$ is the distribution of genotype $a$ and $<\overline{x}>$ as the average of $\overline{x(a)}$ over all genotypes.

Now again, assuming the Gaussian distribution, the distribution $P(x,a)$ is written as follows:

\begin{equation}
P(x,a)= \widehat{N}\exp[ -\frac{(x-X_0)^2}{2\alpha(a)}+\frac{C(a-a_0)(x-X_0)}{\alpha}-\frac{1}{2\mu}(a-a_0)^2],
\end{equation}

\noindent
With $\widehat{N}$ as a normalization constant. The Gaussian distribution $\exp(-\frac{1}{2\mu}(a-a_0)^2)$ represents the distribution of genotypes around $a=a_0$, whose variance is (in a suitable unit) the mutation rate $\mu$. The coupling term $c(x-X_0)(a-a_0)$ represents the change of the phenotype $x$ by the change of genotype $a$, so that
the average phenotype value for given genotype $a$ satisfies
\begin{equation} 
\overline{x}_a \equiv \int x P(x,a) dx=X_0+C(a-a_0).
\end{equation}

Now, the above distribution (5)  can then be rewritten as:

\begin{equation}
P(x,a)= \widehat{N}\exp[-\frac{(x-X_0-C(a-a_0))^2}{2\alpha(a)}+(\frac{C^2}{2\alpha(a)}-\frac{1}{2\mu})(a-a_0)^2].
\end{equation}

Now we postulate {\bf evolutionary stability}, i.e.,  at each stage of the evolutionary course, the distribution has a single peak in $(x,a)$ space. This assumption is rather natural for evolution to progress robustly and gradually. When a certain region of phenotype, say $x> x_{thr}$, is selected, the evolution to increase $x$ works if the distribution is concentrated. On the other hand, if the peak is lost and the distribution is flattened, then selection to choose a fraction of population to increase $x$ no longer works, as the distribution is extended to very low-fitness values. (see also the remark at the end of this section). In the above form, this stability condition is given by

\begin{math}
\frac{C^2}{2\alpha}-\frac{1}{2\mu} \leq 0, i.e.,
\end{math}
\begin{equation}
\mu \leq \frac{\alpha}{C^2} \equiv \mu_{max}.
\end{equation}

\noindent
This means that the mutation rate has an upper bound $\mu_{max}$ beyond which the distribution does not have a peak in the genotype--phenotype space. Beyond this mutation rate, the distribution is extended to very low values of $x$ (fitness). A remark should be done here: In some other form of the distribution (say $exp\{-(x-a)^2/(2\mu \}))$ i.e., with a different coupling form between  $x$ and $a$ than eq.(5), the stability could always hold.  In other words, we assumed the existence of such loss of stability with the increase of mutation rate here. 
There is a transition to to produce mutants with very low fitness values.

Such loss of stability of a fitness distribution was also discussed as error catastrophe by Eigen and Schuster(1979),
where the accumulation of nonfit mutants occur by their combinatorial variety with the increase of the mutation rate.
In contrast to their case without stochasticity in the genotype-phenotype mapping,
the present error catastrophe occurs also with the {\sl decrease} of noise in developmental process, 
following the instability in the genotype-phenotype mapping itself
(via instability in developmental dynamics, as will be shown in the next section).

Note that $\alpha$ is the phenotypic variance $<\delta x^2>$ of isogenic individuals, $V_{ip}$ for the same genotype $a_0$. Hence the inequality is rewritten as

\begin{math}
V_{ip}(a_0) \geq \mu C^2 \end{math}. 
Now, recalling $V_g=<(\overline{x}_a -X_0)^2>$ and eq.(6), $V_g$ is given by $C^2<(\delta a)^2>$. Here, $(\delta a)^2$ is computed by the average $P(x,a)$, so that it is obtained by $\widehat{N}\int  (a-a_0)^2 \exp[(\frac{C^2} {2\alpha(a)}-\frac{1}{2\mu})(a-a_0)^2]$. From this formula we obtain
\begin{equation} V_g= \frac{\mu C^2}{1-\mu C^2/\alpha}. \end{equation} 
Similarly, we obtain the average $V_{ip}$ over all genotypes $a$ from the distribution to get
\begin{equation}
\overline{V_{ip}}=\frac{\alpha}{1-\mu C^2 /\alpha}.
\end{equation}  Hence the inequality (8) is rewritten  as:
\begin{equation}
V_{g} \leq \overline{V_{ip}}.
\end{equation}

Now, by recalling the definition of the $\mu_{max}=\alpha/C^2$ where the inequality is replaced by equality, and by using 
$\overline{V_{ip}}/V_g=\alpha/(\mu C^2)$, we get
\begin{math}
V_{g}=\frac{\mu}{\mu_{max}} \overline{V_{ip}}.
\end{math}
As long as  the mutation rate and $\mu_{max}$ do not change through the course of evolution,  the proportionality between $\overline{V_{ip}}$ and $V_g$ are derived.
To sum up the present formulation of distribution,  we have derived (Kaneko and Furusawa 2006, 2009, Kaneko 2006)

(i) $\overline{V_{ip}}\ \geq V_{g}$, where $\overline{V_{ip}}$ is the average of
$V_{ip}$ over individuals with different genotypes, and 

(ii) proportionality between $V_g$ and $\overline{V_{ip}}$ 
through a given course of evolution. 

Note that there are three assumptions in the ``derivation". First, we assumed existence of a two-variable distributions in genotype and phenotype $P(x,a)$. 
Whether genotype is represented by a continuous parameter is not a trivial assumption. Second, the stability assumption is 
introduced for robust and gradual evolution, postulating that $P(x,a)$ has a single peak in $(x,a)$ space at each generation. Third,  existence of threshold mutation rate $\mu_{max}$ is assumed, beyond which the stability condition collapses. In other words, existence of error catastrophe is implicitly assumed.
With these three assumptions, the error catastrophe is shown to occur at $V_{g} \approx \overline{V_{ip}}$.
To derive the proportionality (ii), additional assumption is made on the constancy of $\mu$ and $\mu_{max}$ through the evolutionary course. 

{\sl Remark}: The argument on the collapse of distribution as $V_g$ approaches $\overline{V_{ip}}$ cannot tell about the distribution form after the collapse.  In the formulation with Gaussian distribution,
the distribution of phenotype is symmetric around the peak. 
However, the fitted state is rare in configurations of genes, in general, and thus the distribution after the error catastrophe is extended to
the lower fitness side.  This discrepancy comes from the choice of one scalar variable for $x$ and $a$.  The configuration space
of genes is very high-dimensional originally, but by focusing on a specific phenotype, the one-dimensional scalar variable is chosen.  Such 'projection' onto one-dimensional space would be valid, a long as the selection of a fitted phenotype works, whereas
near the collapse of the distribution, it is no longer valid, and one needs to consider high-dimensional configuration space of
genes. Then, after the collapse of single-peaked-ness, the distribution is extended to lower fitness only, because the configuration of genes leading to lower fitness is much larger.
Accordingly, it is expected that the evolution to select a higher fitness phenotype no longer works for  $V_{g} \geq \overline{V_{ip}}$.

\begin{figure}[tbp]
\begin{center}
\includegraphics[width=12cm,height=8cm]{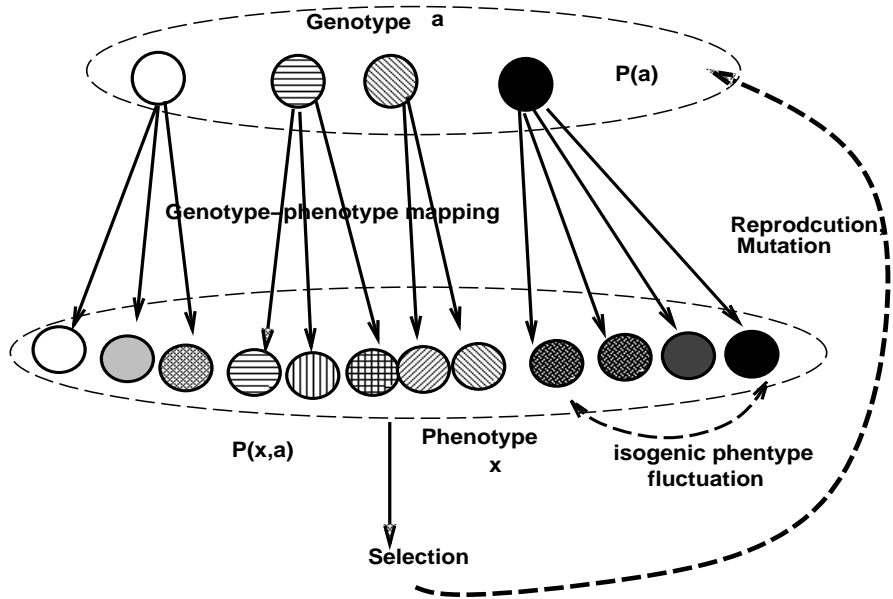}
\caption{
Schematic representation of genotype-phenotype relationship.
The genotype $a$ is distributed by $p(a)$, whereas there is a stochastic mapping from the
genotype to phenotype as a result of developmental dynamics with noise.
Thus the phenotype $x$ is distributed even for isogenic individuals, so that the
distribution $P(x,a)$ is defined.  Mutation-selection process works on phenotype $x$, so that
a feedback from $P(x,a)$ to the genotype distribution $p(a)$ is generated.}
\end{center} 
\end{figure}

\section{Microscopic Approach:  Gene Regulation Network Model}

It is also necessary to study the evolution of robustness and phenotypic fluctuations ``microscopically", by considering ``developmental" dynamics of phenotype.  To introduce such dynamical systems model, one needs to consider the following structure of evolutionary processes.

(i) There is a population of organisms with a distribution of genotypes.

(ii) Phenotype is determined by genotype, through ``developmental" dynamics.

(iii) The fitness for selection is given by the phenotype.

(iv) The distribution of genotypes for the next generation is a result of reproduction, mutation, and selection.  Mutations change the genotype, whereas the offspring number will be based on the fitness value.

During evolution {\sl in silico}, procedures (i), (iii), and (iv) are adopted as genetic algorithms. However, it is important to note that the phenotype is determined only after complex `developmental' dynamics (ii), which are stochastic due to the noise therein.

We have studied gene expression dynamics that are governed by regulatory networks(Glass and Kauffman 1973, Myolsness et al., 1991, Salzar-Ciudad et al., 2000).  Each expression profile changes in time, and the ``developed"
pattern determines evolutionary fitness. Selection occurs after the introduction of mutation at each generation in the gene network.  Among the mutated networks, we select those networks with higher fitness values.  
The dynamics of a given gene expression
level $x_i$ is described by:

\begin{equation}
dx_i/dt =f(\sum_{j}^{M} J_{ij}x_j)-x_i +\sigma \eta _i(t),
\end{equation}

\noindent
where $J_{ij}=-1,1,0$, and $\eta_i(t)$ is Gaussian white noise given by$<\eta _i(t)\eta _j(t')>= \delta _{i,j}\delta(t-t')$.  $M$ is the total number of genes. The function $f$ represents gene expression dynamics following Hill function, or described by a certain threshold dynamics(Kaneko 2007,2008).

The value of $\sigma$ represents the noise strength
that determines stochasticity in gene expression.
We prefix the initial condition for this ``developmental" dynamical system (say, with the condition that no genes are expressed). The phenotype is a pattern of $\{x_i\}$ developed after a certain time.


The fitness $F$ is determined as a function of $x_i$.  Sometimes, not all genes contribute to the fitness.  It is given a function of only a set of ``output" gene expressions.  In the paper(Kaneko,2007), 
we adopted the fitness as the number of expressed genes among the output genes $j=1,2,\cdots,k<M$, i.e., $x_j$ grater than a certain threshold.
Selection is applied after the introduction of mutation at each generation in the transcriptional regulation network, i.e., the matrix $J_{ij}$.

Among the mutated networks, we select a certain fraction of networks that has higher fitness values.
Because the network is governed by $J_{ij}$ which determines the `rule' of the dynamics,
it is natural to treat $J_{ij}$ as a measure of genotype.
Individuals with different genotype have a different set of $J_{ij}$.

As the model contains a noise term, whose strength is given by $\sigma$, the fitness can fluctuate among individuals sharing the same genotype $J_{ij}$. Hence the fitness $F$ is distributed.
This leads to the isogenic phenotype variance
denoted by $V_{ip}(\{J\})$ for a given genotype (network)  $\{J\}$  i.e.,

\begin{equation}
\overline{V_{ip}}=\int d\{J\} p(\{J\}) \int dF \hat{P}(F;\{J\}) (F-\overline{F}(\{J\}))^2 ,
\end{equation}

\noindent
where $\hat{P}(F;\{J\})$ is the fitness distribution over isogenic species  sharing the same network $J_{ij}$, and
$\overline{F}(\{J\})=\int F \hat{P}(F;\{J\}) dF$ is the average fitness, and
$ p(\{J\})$ is the population distribution of the genotype $\{J\}$.
At each generation, there exists a population of individuals with different genotypes  $\{J\}$. The phenotypic variance by genetic distribution is given by

\begin{equation}
V_g=\int d\{J\} p(\{J\}) (\overline{F}(J)-<\overline{F}>)^2 ,
\end{equation}

\noindent
where  $<\overline{F}>= \int p(\{J\})\overline{F}(\{J\}) d\{J\}$ is
the average of average fitness over all populations.

Results from several evolutionary simulations of this class of models are summarized as follows:

(i) There is a certain critical noise level $\sigma_c$, beyond which the evolution of robustness progresses, so that both $V_g$ and $\overline{V_{ip}}$ decrease.  All the phenotypes approach the fittest state with decreasing the variance.  For a lower level of noise $\sigma < \sigma_c$, mutants that have very low fitness values always remain, and evolution does not progress to further increase the average fitness.

(ii) Around the transition point of the noise level,
$V_g \sim \overline{V_{ip}}$, whereas for $\sigma>\sigma_c$, $\overline{V_{ip}}>V_g$. For robust evolution to progress, this  inequality is satisfied.

(iii) When noise is larger than this threshold, through the evolution course$V_g \propto \overline{V_{ip}}$ holds (see Fig.3).

Thus, all the theoretical predictions are supported by the simulations.

Why does the system not maintain the highest fitness state under small
phenotypic noise $\sigma<\sigma_c$?  Indeed, the dynamics of the top-fitness networks that evolved
under low noise ($\sigma < \sigma_c$) have distinguishable dynamics from those that evolved under high noise. It was found that,
for networks evolved under $\sigma>\sigma_c$, a large portion of the initial
conditions reach attractors that give higher fitness values, while for those evolved under
$\sigma<\sigma_c$, only a tiny fraction (i.e., the vicinity of all-off states) is attracted to them.

When the time course of a given gene expression pattern to reach its final pattern
is represented as a motion falling along a potential valley, our results suggest
that the potential landscape becomes smoother and simpler through evolution and
loses ruggedness after a few hundred generations.
Only under
$\sigma>\sigma_c$, the `developmental' landscape as proposed by Waddington evolves to decrease ruggedness and to form global smooth attraction (see Fig.4). In fact, this type of developmental dynamics with global and smooth attraction is know in protein folding dynamics (Abe and Go, 1980, Onuchic et al., 1995), gene expression dynamics(Li et al., 2004), and signal transduction network(Wang et al. 2006).

Now, consider mutation to a network to change one or a few elements of the matrix $J$.  The change in the network leads to slight alterations in gene expression dynamics.
In smooth landscape with global basin of attraction,
this perturbation influences the
attraction to the target only slightly. In contrast, under the dynamics with rugged developmental landscape, a slight change easily destroys the attraction to the
target attractor.  For this latter case, very low-fitness mutants appear, which explains the behavior discussed so far.
In other words, evolution to eliminate
ruggedness in developmental potential is possible only for sufficient noise amplitude.

We expect these relationships (i)-(iii) as well as the existence of the error catastrophe to be generally valid for systems satisfying the following conditions.

(A) Fitness is determined through developmental dynamics.

(B) Developmental dynamics is complex so that its orbit, when deviated by noise, may fail to reach the state with the highest fitness.

(C) There is effective equivalence between mutation and noise in the developmental dynamics with regards to phenotype change.

Condition (A) is generally satisfied for any biological system.  Even in a unicellular organism, phenotype is shaped through
gene expression dynamics.  The condition (B) is the assumption on the existence of 'error catastrophe'.  A phenotype state with a
higher function is rather rare, and to shape such phenotype, the dynamics involve many degrees of freedom and suitable selection of
developmental path.  In our model, it is satisfied because of the complex gene expression dynamics to reach a target expression pattern.
In a system with (A)-(B), slight differences in genotype may lead to huge differences in phenotype, as the phenotype is determined 
after temporal integration of the developmental dynamics where slight differences in genotype (i.e., in the rule that governs the dynamics) are accumulated and amplified.
This is the origin of ``error catastrophe".
Postulate (C) also seems to be satisfied generally; by noise, developmental dynamics are modulated to change the path in the developmental course,
whereas some change in some terms in the equation governing the developmental dynamics caused by mutation can have a similar effect.
In the present model, the phenotypic fluctuation to increase synthesis of a specific gene product
or mutation to add another term to increase such production can contribute to the increase in the expression level of a given gene in the same manner. 

\begin{figure}[tbp]
\begin{center}
\includegraphics[width=12cm,height=9cm]{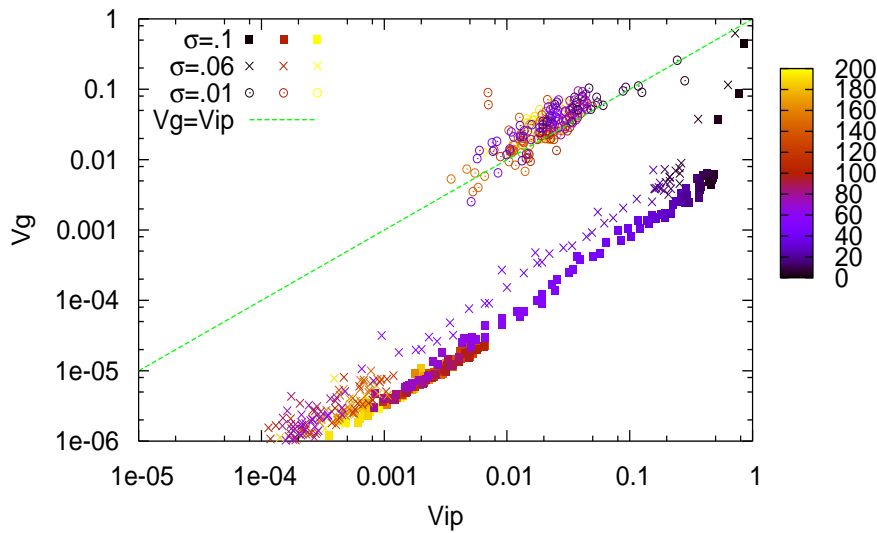}
\caption{ The relationship between $V_g$ and $\overline{V_{ip}}$.  $V_g$ is computed from
the distribution of fitness over different $\{J\}$
at each generation(eq.(14) , and $\overline{V_{ip}}$ by averaging the
variance of isogenic phenotype fluctuations over all existing individuals(eq.(13). Computed by
using the model eq.(12) ( see (Kaneko,2007)  for detailed setup).
Points are plotted over 200 generations, with gradual color change as displayed.
$\sigma=.01$ ($\circ$), .06 ($\times$), and .01 ($\Box$).
For $\sigma> \sigma_c\approx .02$, both decrease with successive generations.}
\end{center}
\end{figure}

\begin{figure}[tbp]
\begin{center}
\includegraphics[width=10cm,height=8cm]{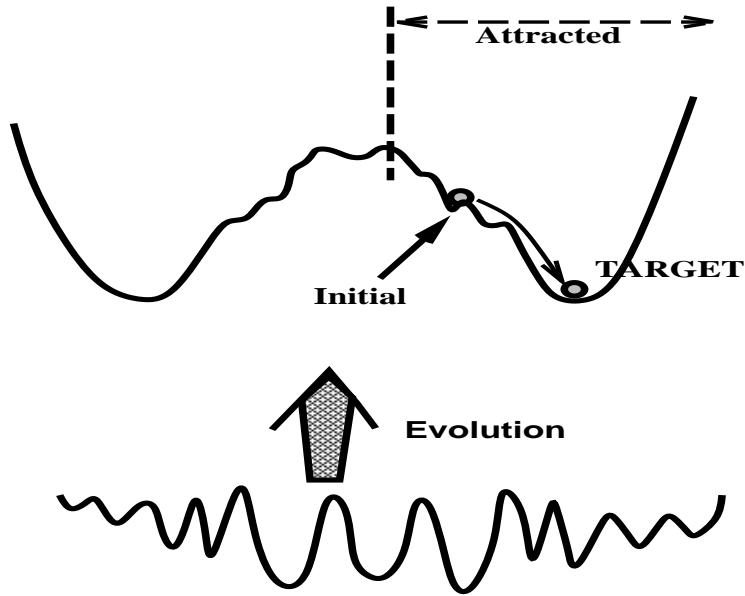}
\caption{ Schematic representation of the basin structure, represented as a
process of descending a potential landscape.  In general, developmental dynamics can have many attractors, and
depending on initial conditions, they may reach different phenotypes, so that the
developmental landscape is rather rugged.   After evolution, global and smooth
attraction to the target phenotype
is shaped by developmental dynamics, if gene expression dynamics are sufficiently noisy. 
}
\end{center}
\end{figure}

\section{Discussion}

Our theory has several implications in development and evolution. We briefly discuss a few topics.

\subsection{Evolvability  and its relationship with fluctuations: answer to Question 1}

In \S 2, we have shown the proportionality between phenotypic plasticity and phenotypic fluctuation $V_{ip}$ as a natural consequence of fluctuation-response relationship.  Then we have shown the proportionality between this epigenetic phenotypic fluctuation $\overline{V_{ip}}$ and genetic variance of phenotype $V_g$ which is  proportional to evolution speed according to Fisher's theorem.  Hence the phenotypic plasticity and evolvability are related through the phenotypic fluctuation.  Although such relationship was discussed qualitatively over decades, quantitative conclusion has not been drawn yet.  Here we have presented such quantitative relationship by distribution theory,  numerical and in-vivo experiments.

Waddington proposed genetic assimilation in which phenotypic change due to the environmental change is later embedded into genes(Waddington 1953,1957).  The proportionality among phenotypic plasticity, $\overline{V_{ip}}$, and $V_g$ is regarded as a quantitative expression of genetic assimilation, since $V_g$ gives an index for evolution speed.

Evolvability may depend on species.  Some species called living fossil preserve the over much longer generations than many others.  A possible answer to such difference in evolvability could be given by the degree of rigidness in developmental process.  If the phenotype generated by developmental process is rigid, then the phenotypic fluctuation as well as phenotypic plasticity against environmental change is smaller.  Then according to the theory here, the evolution speed is reduced.

Evolvability depends not only on species but on each gene in a given species. Some protein could evolve more rapidly than others. In our theoretical study (Kaneko 2008), correlation between phenotypic fluctuation level and evolution speed is confirmed not only over different individuals or species but also for different genes in a given species.  Consider again the model discussed in \S 4, consisting of expression levels of a variety of genes. Recently we have found the proportionality between $V_g$ and $\overline{V_{ip}}$ over expressions of a variety of  genes.  We have measured the variances for expression of each gene $i$, and defined their genetic and epigenetic variances $V_g(i)$ and $\overline{V_{ip}(i)}$ for each.  From several simulations, we have confirmed the proportionality between $V_g(i)$ and $\overline{V_{ip}(i)}$ over many genes $i$. Although direct experimental support is not available yet, recent data on isogenic phenotypic variance and mutational variance over proteins in yeast may suggest the existence of such correlation between the two(Landry et al. 2007, Lehner 2008).

\subsection{Evolution of robustness and relevance of noise: Answer to the Question 2}

As mentioned in the introduction, robustness is insensitivity of fitness to system's changes.  Thus, the  developmental robustness to noise is characterized by the inverse of $\overline{V_{ip}}$, while the mutational robustness is  by the inverse of $V_g$.  In  \S 4, it is shown that the two variances decrease in proportion through evolution under a fixed fitness condition if there exits a sufficient level of noise in the development.  This means the evolution of robustness, as was proposed as Shmalhausen's stabilizing selection(Schmahausen 1949) and Waddington's canalization(Waddington 1942,1957). 

On the other hand, the proportionality between the two variances mean the proportionality between developmental and mutational robustness. Note that this evolution of robustness is possible only under sufficient level of noise in development.  Robustness to noise in development brings about robustness to mutation.

The result demonstrates the role of noise during phenotype expression dynamics to evolution of robustness. Despite recent quantitative observations on phenotypic fluctuation,
noise is often thought to be an obstacle in tuning a system to achieve and maintain a state with higher functions, 
because the phenotype may be deviated from an optimal state achieving a higher function through such noise. Indeed, the question most often 
asked is how a given biological function can be maintained in spite of the existence of phenotypic noise(Kaern et al 2005, Ueda et al., 2001). Here positive role of noise to evolution was revealed quantitatively, whereas relevance of noise to adaptation(Kashiwagi et al., 2006, Furusawa and Kaneko 2008) and to cell differentiation(Kaneko and Yomo, 1999) was discussed earlier.

In our theory, robustness in genotype and phenotype are represented by the single-peaked-ness in the distribution $P(x=phenotype,a=genotype)$.  If we write the distribution in a ``potential form"
as
\begin{equation}
P(x,a)=exp(-U(x,a))
\end{equation}
stability condition means existence of global minimum of the potential $U(x,a)$. With the increase in the slope of the attraction in the potential, the robustness increases.  The smooth attraction in the developmental potential shown in \S 4 is a representation of canalization in epigenetic landscape by Waddington, whereas the correspondence in the attractions both in phenotypic ($x$) and genotypic ($a$) is one representation of genetic assimilation.

\subsection{Future issues}

All of plasticity, fluctuation, and evolvability decrease in our model simulations through the course of evolution, under selection by fixed fitness condition.  Considering that non of the phenotypic plasticity, fluctuation, and evolvability vanish in the present organisms in nature, some process for restoration or maintenance of plasticity or fluctuation should also exist.  To consider such process, the followings are necessary:

1. Fitness with several conditions:  In our model study (and also in experiments with artificial selection), fitness condition is imposed by a single condition.  In nature, the fitness may consist in several conditions.  In this case, we may need more variables in characterization of the distribution function $P$, say $P(x,y,a,b)$ etc.  In such multi-dimensional theory, interference among 'phenotypes' $x$ and $y$ may contribute to the increase or maintenance of fluctuations.
2.

2. Environmental variation: If the environmental condition changes in time, the fluctuation level may be sustained.  To mimic the environmental change, we have altered the condition of target genes to be expressed at some generation in our model so that some target genes should off instead of on.  After this alteration, both the fluctuations $V_{ip}$ and $V_g$ increase over some generations, and then decrease to fix the expression of target genes.  It is interesting that both the expression variances increase keeping proportionality.  At any rate, the restoration of fluctuation levels appears as a result of environmental variation. After the environment changes, both the variances increase to adapt the new environmental condition.  The variances again work as a measure of plasticity. It is further necessary to incorporate environmental fluctuation into our theoretical formulation.

3. Interaction:
Another source to sustain the level of fluctuation is interaction among individuals, which introduces a change in the `environmental condition' of each individual faces.  Here as the variation in phenotypes is larger, the variation in interaction will be increased. Due to the variation in interaction term, the phenotypic variances will be increased. Then,  mutual reinforcement between diversity in interaction and phenotypic plasticity is expected, so that both the genetic and phenotypic diversity are resulted.  Coevolution of phenotypic plasticity and genetic diversity sill be an important issue to be studied in future.

4. Speciation:
So far, we have considered gradual evolution keeping a single peak in phenotype distribution.  On the other hand, the peak splits into two (or more) at the speciation.   Previously we have shown that robust sympatric speciation can occur as a mutual reinforcement of phenotypes bifurcated as a result of developmental dynamics(Kaneko and Yomo, 2000, Kaneko 2002).  It would be important to reformulate the problem in terms of the present distribution theory.

5. Sexual reproduction:
Our simulation in the present paper was based on asexual haploid population. On the other hand, our stability theory on the distribution could be basically applicable also to a population with sexual reproduction.  However, in the sexual reproduction, recombination is a major source of genetic variation besides the mutation.  Then instead of error catastrophe by mutation, the recombination rate could be a source for error catastrophe.  However, the degree of genetic heterogeneity induced by recombination strongly depends on the genotype distribution at the moment.  (For population of clones, the mutation can produce heterogeneity, but the recombination only cannot produce the heterogeneity). It would be interesting to extend our theory on evolutionary stability and phenotypic fluctuation to incorporate the sexual reproduction.

\subsection{Conclusion}

We have demonstrated general relationship among phenotypic plasticity, genetic and epigenetic fluctuation of phenotype, and evolution speed.  Increase of developmental and mutational robustness through evolution is demonstrated, which progresses keeping the proportionality between the two.

Recall  Waddington' canalization, epigenetic landscape, and genetic assimilation (Waddington, 1957). Our distribution function $P(phenotype=x,genotype=a) =exp(-U(x,a)$ and its stability give one representation of the epigenetic landscape. In terms of fluctuations, the canalization and genetic assimilation are formulated quantitatively. Note that the existence of the phenotype-genotype distribution is based on mutual relationship between phenotype and genotype: genotype determines (probabilistically) phenotype through developmental dynamics, while phenotype restricts genotype through mutation-selection process.  For robust evolutionary process the consistency between phenotype and genotype are assumed to be formed which leads to the $P(phenotype,genotype)$ assumptions, and the general relationships proposed here.

Recall that such ``consistency principle" was adopted by Einstein to formulate fluctuation-dissipation relationship(Einstein, 1926), where consistency between microscopic and macroscopic descriptions was a guiding principle for the formulation. We hope that quantitative formulation of Waddington's ideas based on Eisntein's spirit will provide a coherent understanding in evolution-development relationship.

\vspace{.1cm}

{\bf Acknowledgement}

I would like to thank C. Furusawa, T. Yomo, K. Sato, M. Tachikawa, S. Ishihara, S. Sawai for continual discussions, E. Jablonka, T. Sato, S Jain,  S.Newman, and V. Nanjundiah for stimulating discussions during the conference at Trivandrum.  Last but not least, I would like to express sincere gratitude to S. Newman and V Nanjundiah for beautiful organization of the stimulating conference.

\vspace{.1cm}

{\bf References}

Abe H. and Go. N., (1980) Noninteracting local-structure model of
folding and unfolding transition in globular proteins. I.
Formulation.,Biopolymers 20, 1013.

Alon U., Surette M.G., Barkai N., and Leibler S.: 1999
`Robustness in bacterial chemotaxis',
{\sl Nature}, {\bf 397}, pp. 168--171

Ancel, L.W., and Fontana, W.: 2002,
`Plasticity, evolvability, and modularity in RNA',
{\sl J. Exp. Zool.}, {\bf 288}, pp. 242--283

Bar-Even A., et al.: 2006,
`Noise in protein expression scales with natural protein abundance',
{\sl Nature Genetics}, {\bf 38}, pp. 636--643

Barkai N., and Leibler S.: 1997,
`Robustness in simple biochemical networks',
{\sl Nature}, {\bf 387}, pp. 913--917

Callahan, H.S., Pigliucci, M., and Schlichting, C.D.: 1997,
`Developmental phenotypic plasticity: where ecology and evolution meet molecular biology',
{\sl Bioessays}, {\bf 19}, pp. 519--525 

Ciliberti S., Martin O.C., and Wagner A.: 2007,
`Robustness can evolve gradually in complex regulatory gene networks with varying topology',
{\sl PLoS Comp. Biology}, {\bf 3}, e15

Eigen M, Schuster P (1979) The Hypercycle, Springer, Heidelberg.

Einstein A., 1926, Investigation on the Theory of
of Brownian Movement, (Collection of papers ed. by R. Furth),
Dover, (reprinted, 1956).

Elowitz, M.B., Levine, A.J., Siggia, E.D., and Swain, P.S.: 2002
`Stochastic gene expression in a single cell',
{\sl Science}, {\bf 297} pp. 1183--1187

Fisher, R.A., 1930, `The Genetical Theory of Natural Selection',
Oxford Univ. Press., (reprinted, 1958)

Furusawa C. and Kaneko K., 2008, ``A generic mechanism for adaptive growth rate regulation", PLoS Computational Biology, {\bf4}, e3

Furusawa, C., Suzuki, T., Kashiwagi, A., Yomo, T., and Kaneko, K.: 2005,
`Ubiquity of log-normal distributions in intra-cellular reaction dynamics',
{\sl Biophysics}, {\bf 1}, pp. 25--31

Gibson G., Wagner G.P., 2000,
Canalization in evolutionary genetics: a stabilizing theory?
 Bioessays 22, 372 -- 380

Glass, L., and Kauffman, S.A.: 1973,
`The logical analysis of continuous, non-linear biochemical control networks',
{\sl J. Theor. Biol.}, {\bf 39}, pp. 103--129

Hasty, J., Pradines, J., Dolnik, M., and Collins, J.J.: 2000,
`Noise-based switches and amplifiers for gene expression',
{\sl Proc. Natl. Acad. Sci. U S A}, {\bf 97}, pp. 2075--2080

Kaern, M., Elston, T.C., Blake, W.J., and Collins, J.J.: 2005,
`Stochasticity in gene expression: from theories to phenotypes',
{\sl Nat. Rev. Genet.}, {\bf 6}, pp. 451--464

Kaneko, K.: 2006, {\sl Life: An Introduction to Complex Systems Biology}, Springer, Heidelberg and New York

Kaneko K.: 2007,
`Evolution of robustness to noise and mutation in gene expression dynamics',
{\sl PLoS ONE}, {\bf 2}, e434

Kaneko K.: 2008,
`Shaping Robust system through evolution',
{\sl Chaos}, {\bf 18}, 026112

Kaneko, K. and Furusawa, C.: 2006,
`An evolutionary relationship between genetic variation and phenotypic fluctuation',
{\sl J. Theo. Biol.},  {\bf 240}, pp. 78--86

Kaneko K. and Furusawa C., 2008, 
Consistency Principle in Biological Dynamical Systems
Theory in Bioscience, in press 

Kaneko, K. and Yomo, T.: 1999,
`Isologous diversification for robust development of cell society',
{\sl J. Theor. Biol.}, {\bf 199}, pp. 243--256

Kaneko K., and Yomo T., 2000,
``Sympatric Speciation: compliance with
phenotype diversification from a single genotype"
Proc. Roy. Soc. B, 267, 2367-2373

Kaneko K., 2002,
``Symbiotic Sympatric Speciation:
Compliance with Interaction-driven Phenotype
Differentiation from a Single Genotype",
Population Ecology, 44, 71-85

Kashiwagi, A., Urabe, I., Kaneko., K., and Yomo, T.: 2006,
`Adaptive response of a gene network to environmental changes by attractor selection',
{\sl PLoS ONE}, {\bf 1}, p. e49

Kirschner, M.W., and Gerhart J.C.: 2005,
`The Plausibility of Life', Yale Univ. Press

Krishna, S., Banerjee, B., Ramakrishnan, T. V.,  and Shivashankar, G. V., 2005,
``Stochastic simulations of the origins and implications of long-tailed
distributions in gene expression", Proc. Nat. Acad. Sci. USA 102,
4771-4776

Kubo R., Toda M., Hashitsume N. 1985,
{\sl Statistical Physics II}: (English translation;  Springer).

Landry C. et al., 2007, ``Genetic Properties Influencing the Evolvability of Gene Expression'', Science, {\bf 317}, 118 - 121

Lehner B., 2008,
``Selection to minimize noise in living systems and its implications for the evolution of gene expression'', Mol Syst Biol. ;{\bf 4}:.170

Li, F., Long, T., Lu, Y., Ouyang, Q., and Tang, C.: 2004,
`The yeast cell-cycle network is robustly designed',
{\sl Proc. Nat. Acad. Sci. U S A}, {\bf 101}, pp. 10040--10046

McAdams H.H., and Arkin A.,1997,
``Stochastic mechanisms in gene expression",
Proc. Natl. Acad. Sci. USA,
{\bf 94}, pp. 814-819  

Mjolsness, E., Sharp, D.H., and Reisnitz, J., 1991
`A connectionist model of development',
{\sl J. Theor. Biol.}, {\bf 152}, 429--453

Onuchic J.N., Wolynes P.G., Luthey-Schulten Z., and Socci
N.D., 1995, ``Toward an Outline of the Topography of a Realistic
Protein-Folding Funnel", Proc. Nat. Acad. Sci. USA 92, 3626.

Oosawa F., 1975,
``Effect of the field fluctuation on a macromolecular system",
J. Theor. Biol. 52, 175.

Pigliucci M.,  Murren C.J., and Schlichting C.D., 2006,
``Phenotypic plasticity and evolution by genetic assimilation",
Journal of Experimental Biology 209, 2362-2367.


Salazar-Ciudad, I., Garcia-Fernandez, J., and Sole, R.V., 2000,
`Gene networks capable of pattern formation: from induction to reaction--diffusion',
{\sl J. Theor. Biol.}, {\bf 205}, 587--603

Sato, K., Ito, Y., Yomo, T., and Kaneko, K., 2003
`On the relation between fluctuation and response in biological systems',
{\sl Proc. Nat. Acad. Sci. U S A}, {\bf 100}, pp. 14086--14090

Schmalhausen, I.I., 1949,
`Factors of Evolution: The Theory of Stabilizing Selection',
Univ. of Chicago Press, Chicago, (reprinted 1986).

Spudich J.L., and Koshland, D.E., Jr.,1976,
`Non-genetic individuality: chance in the single cell',
{\sl Nature}, {\bf 262}, pp. 467--471

Ueda, M., Sako, Y., Tanaka, T., Devreotes, P., and Yanagida, T., 2001,
`Single-molecule analysis of chemotactic signaling in {\it Dictyostelium} cells',
{\sl Science},  {\bf 294}, pp. 864--867

de Visser J.A., et al., 2003,
`Evolution and detection of genetic robustness',
{\sl Evolution}, {\bf 57}, pp. 1959--1972.

Waddington, C. H. ,1942, `` Canalization of Development and the
Inheritance of Acquired Characters", Nature {\bf 150},563--565

Waddington C.H.,1953, ``Genetic Assimilation of an Acquired Character",
Evolution, 7, 118--126

Waddington, C.H., 1957,
{\sl The Strategy of the Genes}, Allen \& Unwin, London

Wagner A.,2000,
`Robustness against mutations in genetic networks of yeast',
{\sl Nature Genetics}, {\bf 24}, pp. 355--361

Wagner A, 2005, {\sl Robustness and evolvability in living systems},
Princeton University Press, Princeton NJ.

Wagner G.P., Booth G., Bagheri-Chaichian H., 1997,
A population Genetic Theory of Canalization,
Evolution {\bf 51} 329-347.

Wang J, Huang B, Xia X, Sun Z., 2006,
Funneled Landscape Leads to Robustness of Cellular Networks: MAPK
Signal Transduction
Biophysical Journal 91:L54-L56

Weinig, C., 2000,
``Plasticity versus Canalization: Population differences in
the timing of shade-avoidance responses",
Evolution {\bf 54} 441-451

West-Eberhard M.J.,2003,
{\sl Developmental Plasticity and Evolution}, Oxford Univ. Press.

\end{document}